\documentclass[a4paper]{article} 

\usepackage[latin1]{inputenc} 
\usepackage[T1]{fontenc} 
\usepackage[english]{babel} 

\ifx\usepackage\undefined
\def\inca#1{\input #1.sty}
\def\incb#1#2{\input #1.sty}
\else 
\newcommand{\inca}[1]{\usepackage{#1}}
\newcommand{\incb}[2]{\usepackage[#2]{#1}}
\fi

\ifx\pdfoutput\undefined
  \incb{graphics}{dvips}
  \incb{epsfig}{dvips}
  \def\dvipsselector#1#2{#1}
  
\else
  \incb{graphics}{pdftex}
  \def\dvipsselector#1#2{#2}
  
\fi

\inca{boxedminipage,url}
\inca{amstext,amsfonts,latexsym}
\ifx\usepackage\undefined\else\inca{latexsym}\fi
\inca{epic}

\def\mc#1#2#3{\multicolumn{#1}{#2}{#3}}

\ifx\frenchname\undefined
\newtheorem{defin}{Definition}
\newtheorem{prop}{Proposition}
\newtheorem{lem}{Lemma}
\newtheorem{theo}{Theorem}
\newtheorem{corol}{Corollary}
\newtheorem{conj}{Conjecture}
\newtheorem{observ}{Observation}
\newtheorem{exemple}{Example}
\else

\newtheorem{theo}{Théorème}

\fi

\def\N{\mathbb N}

\ifx\mathbb\undefined
\def\mathbb{\Bbb}
\fi

\def\ulp{\mathop{\rm ulp}\nolimits}

\newenvironment{pfigure}{\begin{figure}}{\end{figure}}
\def\figdir{Fig} 
\def\figext{\dvipsselector{pstex_t}{pdftex_t}}

\newcommand{\figtex}[2]{%
  \begin{pfigure}
    \begin{center}  
      \input \figdir/#1.\figext
    \end{center}
    \caption{#2}
    \label{fig/#1}
    \label{fig:#1}
  \end{pfigure}
}

\newcommand{\boxlabel}{}
\newcommand{\boxcaption}{}

\usepackage{times,palatino}
\ifx\titleLIP\undefined
  \ifx\RRetitle\undefined
    \newcommand{\RRItitle}[1]{}
    \newcommand{\RRItitre}[1]{}
    \newcommand{\RRIdate}[1]{}
    \newcommand{\RRIauthor}[1]{}
    \newcommand{\RRIthead}[1]{}
    \newcommand{\RRIahead}[1]{}
    \newcommand{\RRIabstract}[1]{\newcommand{\RRIpre}{\begin{abstract}#1\end{abstract}}}
    \newcommand{\RRIresume}[1]{}
    \newcommand{\RRIkeywords}[1]{}
    \newcommand{\RRImotscles}[1]{}
    \newcommand{\RRIbegin}{}
    \newcommand{\RRInumber}[1]{}
    \newcommand{\RRItheme}[1]{}
    \newcommand{\RRIprojet}[1]{}
    \newcommand{\RRIno}[1]{#1}
    \newcommand{\RRIyes}[1]{}
    \newcommand{\RRIinria}[1]{}
    \newcommand{\RRIlip}[1]{}
  \else
    \newcommand{\RRItitle}[1]{\RRetitle{#1}}
    \newcommand{\RRItitre}[1]{\RRtitle{#1}}
    \newcommand{\RRIdate}[1]{\RRdate{#1}}
    \newcommand{\RRIauthor}[1]{\RRauthor{#1}}
    \newcommand{\RRIthead}[1]{\titlehead{#1}}
    \newcommand{\RRIahead}[1]{\authorhead{#1}}
    \newcommand{\RRIabstract}[1]{\RRabstract{#1}}
    \newcommand{\RRIresume}[1]{\RRresume{#1}}
    \newcommand{\RRIkeywords}[1]{\RRkeyword{#1}}
    \newcommand{\RRImotscles}[1]{\RRmotcle{#1}}
    \newcommand{\RRIbegin}{\URRhoneAlpes \makeRR}
    \newcommand{\RRIpre}{}
    \newcommand{\RRInumber}[1]{}
    \newcommand{\RRItheme}[1]{\RRtheme{#1}}
    \newcommand{\RRIprojet}[1]{\RRprojet{#1}}
    \newcommand{\RRIno}[1]{}
    \newcommand{\RRIyes}[1]{#1}
    \newcommand{\RRIinria}[1]{#1}
    \newcommand{\RRIlip}[1]{}

  \fi
\else
    \usepackage{fancyheadings}
    \newcommand{\RRItitle}[1]{\titleLIP{#1}}
    \newcommand{\RRItitre}[1]{}
    \newcommand{\RRIdate}[1]{\dateLIP{#1}}
    \newcommand{\RRIauthor}[1]{\authorLIP{#1}}
    \newcommand{\RRIthead}[1]{\newcommand{\RRItheadv}{#1}}
    \newcommand{\RRIahead}[1]{\newcommand{\RRIaheadv}{#1}}
    \newcommand{\RRIabstract}[1]{\abstractLIP{#1}}
    \newcommand{\RRIresume}[1]{\resumeLIP{#1}}
    \newcommand{\RRIkeywords}[1]{\keywordsLIP{#1}}
    \newcommand{\RRImotscles}[1]{\motsclesLIP{#1}}
    \newcommand{\RRIbegin}{\rapportLIP\pagenumbering{arabic}\pagestyle{fancy}\rhead[\RRIaheadv]{\thepage}\lhead[\thepage]{\RRItheadv}\cfoot{}}
    \newcommand{\RRIpre}{}
    \newcommand{\RRInumber}[1]{\numberLIP{#1}}
    \newcommand{\RRItheme}[1]{}
    \newcommand{\RRIprojet}[1]{}
    \newcommand{\RRIno}[1]{}
    \newcommand{\RRIyes}[1]{#1}
    \newcommand{\RRIinria}[1]{}
    \newcommand{\RRIlip}[1]{#1}
\fi

\usepackage{color,amssymb}

\RRItitle{Computer validated proofs of \\ a toolset for adaptable arithmetic}

\RRItitre{Boîte à outils pour l'arithmétique numérique \\ validée par un preuve sur ordinateur}

\RRIdate{Janvier 2001} 

\RRIauthor{%
  Sylvie Boldo\RRIinria{, LIP} \\
  Marc Daumas\RRIinria{, CNRS} \\
  Claire Moreau-Finot\RRIinria{, LIP} \\
  Laurent Théry\RRIinria{, INRIA \\~}
}%

\RRIthead{Computer validated proofs of a toolset for adaptable arithmetic} 
\RRIahead{Sylvie Boldo, Marc Daumas, Claire Moreau-Finot and Laurent Théry} 

\RRInumber{2001-01} 

\RRIabstract{%
  Most  existing  implementations  of  multiple  precision  arithmetic
  demand  that the  user  sets  the precision  {\em  a priori}.   Some
  libraries  are said  adaptable in  the sense  that  they dynamically
  change the precision of  each intermediate operation individually to
  deliver  the target  accuracy according  to the  actual  inputs.  We
  present in this text a new adaptable numeric core inspired both from
  floating point expansions and from on-line arithmetic.
  
  The numeric core is cut down  to four tools.  The tool that contains
  arithmetic operations is proved to  be correct. The proofs have been
  formally checked  by the Coq  assistant.  Developing the  proofs, we
  have formally proved many results published in the literature and we
  have extended  a few of  them. This work  may let users  (i) develop
  application specific adaptable libraries  based on the toolset and /
  or (ii) write new formal proofs based on the set of validated facts.
}%

\RRIresume{%
  La plupart  des codes à  précision multiple existants  demandent que
  l'utilisateur spécifie la précision des calculs {\em a priori}.  Une
  solution est d'utiliser pour les calculs intermédiaires la précision
  la  plus  élevée  pour  atteindre  la  précision  demandée  pour  le
  résultat. On  dit que certaines bibliothèques  sont adaptatives dans
  le sens  qu'elles changent dynamiquement la précision  de chacun des
  calculs  intermédiaires  en  fonction  des  valeurs  effectives  des
  entrées pour atteindre la précision demandée pour le résultat.  Nous
  présentons  dans  ce  texte  un nouveau  noyau  numérique  adaptatif
  inspiré  à  la  fois  des  expansions de  nombres  flottants  et  de
  l'arithmétique en-ligne.
  
  Le noyau  numérique se décompose  en cinq outils.  Le  premier outil
  qui contient de nombreuses opérations arithmétiques est prouvé.  Les
  preuves  ont  été validées  formellement  par  l'assistant Coq.   En
  développant  les  preuves, nous  avons  redéveloppé formellement  de
  nombreux résultats déjà publiés dans la littérature et nous en avons
  étendu  certains.   Ce travail  permettra  aux  utilisateurs de  (i)
  développer   des  bibliothèques   adaptatives   spécifiques  à   une
  application basées  sur la  boîte à  outils et /  ou (ii)  écrire de
  nouvelles preuves formelles basées sur notre socle de faits validés.
}%

\RRIkeywords{Multiple  precision,  expansion,  on-line, formal  proof, Coq.}

\RRImotscles{Précision multiple, expansion, en-ligne, preuve formelle, Coq.}

\RRItheme{2}

\RRIprojet{Arénaire}

\RRIinria{\RRnote{%
    This  text  is  also  available   as  a  research  report  of  the
    Laboratoire    de     l'Informatique    du    Parallélisme    {\tt
      http://www.ens-lyon.fr/LIP}.
}}%

\begin{document} 

\RRIbegin

\newtheorem{proc}{Lemma}  
\newcommand{\ptys}[2]{{\noindent \bf{#1}}{\it{#2}}}

\title{%
  {\bf
    Computer validated proofs of \\
    a toolset for adaptable arithmetic}\thanks{%
    Authors  may be  reached via  e-mail  at Sylvie.Boldo@ENS-Lyon.Fr,
    Marc.Daumas@ENS-Lyon.Fr,       Claire.Moreau@ENS-Lyon.Fr       and
    Laurent.Thery@INRIA.Fr.  \RRIlip{This text  is also available as a
      research  report  of  the  Institut  National  de  Recherche  en
      Informatique et en Automatique {\tt http://www.inria.fr}.}
    }%
}%

\author{
  {\bf Sylvie Boldo, Marc Daumas, Claire Moreau-Finot} \\
  Laboratoire de l'Informatique du Parallélisme        \\
  UMR 5668 - CNRS - ENS de Lyon - INRIA                \\[24pt]
  {\bf Laurent Théry}                                  \\
  INRIA Nice---Sophia-Antipolis
}%
\date{}

\maketitle
\RRIpre

\section*{Introduction}

We will  first present examples of numerical  computations coming from
four  fields  of application.   With  each  example  we introduce  new
properties, new goals and some needed references.  As the subject will
become clear, we  will present the contributions of  this text and its
summary.

\subsection*{Motivation}

As an undergraduate student, we  have all learnt to compute a quantity
with a few significant digits first  as 5 miles is about 8 kilometers. 
If the question is ``Are 4.995 miles more than 8 kilometers?'', we are
close to  a threshold. To answer  the question, we  would compute more
(less significant) digits  to find that 4.995 miles  are a little over
38  meters   longer  than   8  kilometers.   We   might  first   do  a
multiplication of 5 by 1.6  and then the exact multiplication of 4.995
by 1.609344.  However, we could save that 5 times 1.6 is exactly 8 and
afterward  just multiply  5 by  9  to obtain  45 meters  with a  small
additional  value  that is  less  than  5  meters.  That  answers  the
question since 0.005 miles is less than 10 meters.

Numerical  analysts have long  designed algorithms  that react  to the
accuracy aimed by the  user.  Countless specific algorithms reduce the
error on the final result by iterative refinement in solving algebraic
equations,  linear systems, ODEs,  PDEs and  so on  \cite{Hig96}. Such
methods are  so robust  that the final  solution can be  very accurate
even if the first attempt before any refinement is not close.

All  the  above-mentioned   solutions  assume  that  the  intermediate
arithmetic operations are performed up  to a fixed level of precision. 
This  level,  usually  the  one  given  by  machine  double  precision
arithmetic, is used for all  the operations and is sufficient to store
the result to  the accuracy aimed by the  user.  Recent work presented
in \cite{Hig96,SorTan91,Bar93}  proposes new enhanced  routines if two
levels of precision are available on the machine. The second level can
be obtained  by a careful use  of the tools presented  in this article
\cite{BolDau01}.  All  these algorithms use the  machine arithmetic as
an  error  prone  tool  that  must  be  circumvented  by  mathematical
analysis.


\subsection*{Adaptability {\em versus} multiple precision}

A  numerical  algorithm  is   only  slightly  modified  to  accommodate
adaptable arithmetic or  multiple precision arithmetic.  With multiple
precision arithmetic, each operator is  fired only once and the result
is computed  with the precision  fixed at compile time.   Static error
analysis is  used to guarantee that  the final result is  known to the
accuracy specified by  the user.  Being static, the  error analysis is
often too pessimistic.  As it is readily available, multiple precision
arithmetic is often chosen by users.

When  a code  has  been  modified to  become  adaptable, every  single
operator is started with a very limited precision.  The runtime system
freezes the operator but it is able to resume its work when necessary.
The  program is  not ended  unless the  final result  is known  to the
accuracy specified by the user. The evaluation continues until this is
the  case,  with  the  system  increasing  automatically  the  working
precision of each operator independently.

Coherency  is key  to the  field of  computational geometry  where one
single tiny error  may change a convex hull  to a non-convex, possibly
non-planar,  graph.   Numerical  routines  are  used  to  answer  such
predicates as ``Is  a point in a circle defined by  3 other points?''. 
A Boolean answer cannot be  approximated, it is either right or wrong,
and one single wrong answer might  lead a program to an abstract state
that cannot  occur in Euclidian geometry.  Recent  works have proposed
clever    implementations    of    multiple    precision    arithmetic
\cite{ForVan93,KarLiPecYap99}  to  answer  correctly such  questions.  
They reinforce the existing  set of general purpose multiple precision
libraries \cite{Bre78a,SerVuiHer89,Bai93,Bai95,Gra2Kb}.  Some marginal
work does implement iterative  refinements of the geometric predicates
\cite{AvnBoiDevPreYvi97}  or change  the algorithm  to  work correctly
with limited predicates \cite{BoiPre2K}.

Adaptability is used for our last example. It is the central scheme of
Ziv's paper \cite{Ziv91} where the author implements correctly rounded
elementary functions (circular and exponential).  Correct rounding for
the  floating   point  numbers   is  mathematically  defined   by  two
international      standards      (ANSI-IEEE-ISO~754      and      854
\cite{Ste.81,Ste.87,Gol91,CodKar.84}).  It  is a monotonous projection
from the  real set to the set  of the floating point  numbers.  If all
the  numbers of  a given  interval round  to the  same  floating point
value,  you can  safely  round this  interval  to the  common result.  
Otherwise, you cannot round the interval.

An approximated algorithm such as  the ones used for the evaluation of
the elementary  functions \cite{Mul97}  does not produce  directly the
rounded result  but it  computes an approximated  result and  an error
bound.  This pair defines an interval that contains the exact result.
\begin{itemize}
\item  If the interval  is narrow,  it rounds  to one  single floating
  point  value  that is  the  correctly  rounded  value of  the  exact
  elementary  function.
\item If we discover from  the current approximation that the interval
  contains a  discriminating point, the approximation  must be refined
  to  reduce its  width.  This  process continues  until  a satisfying
  approximation is obtained to round the interval to one single value.
  The set of the discriminating  point is the set of the representable
  numbers for the directed roundings and the set of the midpoints when
  rounding to the nearest.
\end{itemize}

With the first  iteration of Ziv's scheme, machine  precision is used. 
For the  result interval to  get sharper, the approximation  scheme is
only slightly modified  but the target precision is  increased and the
arithmetic  operations become  increasingly slow.  Cases where  a high
precision result  is needed are  very infrequent and the  running time
seen by  a user will most  probably be the  time for the first  or the
second iteration.

Adaptable behaviour  is sometime recreated with Mathematica  or Maple. 
These two softwares let the user change the working precision and they
provide some criteria to estimate the final accuracy of many numerical
routines \cite{ChaGedGonMonWat91}.  Users set  the number of digits of
the intermediate precision after a few educated random attempts.

\subsection*{Prior art}

\paragraph{Adaptability}

Some  few  authors  already  presented  a  general  purpose  adaptable
arithmetic                                                      library
\cite{Boe87,Men94,Men95,BerDau97,MicMor97,BurFleMehSch99},   but  some
others stopped with one or two adaptable attempts followed by an exact
evaluation   \cite{ForVan93,She97}.     These   last   solutions   are
appropriate in many cases because  the first few attempts are fast and
they succeed  by far in the  most frequent case  for many applications
\cite{DevPre98}. These authors where  forced to stop because they used
a  data structure  that  is not  appropriate  for adaptability.   They
forced  one  or  two  rounds  of  adaptability  with  great  efforts.  

\paragraph{Floating point expansion}

This  data structure was  introduced by  Priest \cite{Pri91}  based on
some earlier  operations \cite{Knu69,Knu73}.  Actually,  M{\o}ller and
Dekker were the first in the past to propose such techniques to extend
precision  on  the  floating  point unit  \cite{Mol65a,Mol65b,Dek71}.  
Kahan  and  Pichat  applied  similar  techniques  for  other  purposes
\cite{Kah65,Pic72}.

Fast algorithms, tightly connected  to the machines, are possible with
IEEE-754 compatible commercial floating point units.  Shewchuk was the
first one  to present a working  library available on the  net with an
actual  application  to  computational  geometry  \cite{She96,She97}.  
Assuming that  the floating point unit  is IEEE compatible  gives us a
powerful set of axiomatic properties on floating point operations.  We
have proposed in the past some algorithms and their implementation for
multiplication        and         division        on        expansions
\cite{Dau99a,DauFin98,DauFin99a}.  Developments are still undergone as
new research teams get  interested by this approach \cite{Sei2K}.  Our
contribution forces  us to define a  new kind of  expansions.  We call
them pseudo-expansions.


\paragraph{On-line arithmetic}

We will  present in  this text that  we can  reduce the growth  of the
running time by setting up an appropriate data structure that will not
restart  an  adaptable evaluation  from  scratch  as the  intermediate
precision is modified.

This  data   structure  is   inspired  by  on-line   arithmetic  where
adaptability is natural.  On line  arithmetic was targeted in the past
to   hardware  design   with  small   radices  (typically   2   or  4)
\cite{TriErc77,CorDupHocMul91,LouErc93,DauMulVui94,DauMulTis97} but it
has    proved    to   be    suitable    for   software    applications
\cite{MazMer93,BerDau97}.

\subsection*{Contributions}

The first contribution  of this work is a  set of primitives available
soon  on   the  Internet  through  our   home  web  site\footnote{{\tt
    http://www.ens-lyon.fr/LIP/Arenaire/}.}   and   through  the  {\sc
  netlib} repository\footnote{{\tt  http://www.netlib.org/} among many
  access protocols.}.  A toy example shows that high level programming
with  object  interface  is  too  slow for  the  kernel  of  numerical
softwares \cite{Gog2K}.   We ran the  same FLOPS routines  three times
\cite{Don92,Don99}.  The first run was performed with object automatic
allocation and destruction.  The second run used in-line function calls
and no object allocation.  The final code used explicit code in-lining.
Removing  object programming  increased the  performance of  133\% and
using code  in-lining yielded  another 56\%.  Our  contributed building
blocks are pieces  of code assembled by the  designer according to its
needs.   They present a  straight interface  to get  a good  trade off
between simplicity and speed.

The second  contribution is a  set of validated proofs.   To mechanise
our    proofs,     we    have    been    using     the    Coq    proof
assistant~\cite{HueKahPau97}.   Systems  like Coq  allow  the user  to
define  new objects and  to derive  consequences of  these definitions
formally. The language of Coq  is based on a higher-order logic.  With
such an expressive logic, it  is possible to state properties in their
most  general form.   For example,  universal quantification  has been
used to state  properties that are true for an  arbitrary format or an
arbitrary  rounding  mode.    Proofs  are  built  interactively  using
high-level  tactics. At the  end of  each proof,  Coq records  a proof
object that  contains all  the details of  the derivation  and ensures
that  the  theorem  is   valid.  Theorem  provers  have  already  been
successfully   used   to  mechanically   check   the  correctness   of
floating-point algorithms~\cite{Rus98,Har97a}.

Our    formal    development    is    freely    available    on    the
Internet\footnote{{\tt  http://www-sop.inria.fr/lemme/AOC/coq/}.}  and
it  will  soon  be   available  as  a  Coq  contribution\footnote{{\tt
    http://coq.inria.fr/}.}.  A clickage map of the hierarchy makes it
possible to browse  into the different components.  At  the moment, it
is 22000 line  long and it contains 106 Definitions  and 545 Theorems. 
No proof is presented in detail in this manuscript, but we outline the
main results  used for the  proofs. Since the proofs  are mechanically
validated, the  reader is  sure that the  annoying special  cases have
been checked and that all the conditions of the properties used in the
proof have been met.  \RRIyes{All the Definitions and Theorems of this
  development  are  presented  in  annexes  A through  W.}   For  each
mentioned  theorem  we give  its  name as  it  appears  in the  formal
development and the file where it is proved.

We first present definitions and validated properties for the floating
point numbers,  the expansions  and the pseudo-expansions.   Section 2
presents  the  addition, multiplication  and  division toolset.   This
presentation ends with concluding remarks in the last Section.


\newcommand{\rr}{\mathbb{R}}
\newcommand{\nn}{\mathbb{N}}
\newcommand{\ideal}[1]{<\!\!#1\!\!>}
\newcommand{\lcm}[2]{#1\!\hbox{\^{}}#2}

\newlength{\ppwd}
\newlength{\ppht}
\newcommand{\rpp}[1]{
\settowidth{\ppwd}{#1}
\settoheight{\ppht}{#1}
\raisebox{\ppht}[0pt]{\makebox[0pt][l]{
\hspace{.15\ppwd}{\tiny \tt >}}}#1}
\newcommand\pp{\rpp{+}}
\newcommand{\rppa}[1]{
\settowidth{\ppwd}{#1}
\settoheight{\ppht}{#1}
\raisebox{\ppht}[0pt]{\makebox[0pt][r]{
{\tiny \tt >}\hspace{-5pt}}}#1}
\newcommand\ppa{\rppa{+}}
\newcommand{\limp}{\Rightarrow}
\newcommand{\reduce}{\rightarrow^{\ }_S}
\newcommand{\reducep}{\rightarrow^+_S}
\newcommand{\reduces}{\rightarrow^*_S}
\newcommand{\mon}{M_n}
\newcommand{\lm}{\le_{M_n}}
\newlength{\hsbw}
\newcommand\MSpacing{13pt}
\newenvironment{cboxed}{\begin{flushleft}
\setlength{\hsbw}{\textwidth}
\addtolength{\hsbw}{-\arrayrulewidth}
\addtolength{\hsbw}{-\tabcolsep}
 \begin{tabular}{@{}|c@{}|@{}}\hline 
 \begin{minipage}[b]{\hsbw}
 \vspace*{.06in}
 \begingroup\small\baselineskip\MSpacing}{\endgroup\end{minipage}\\ \hline 
 \end{tabular}
 \end{flushleft}}
\newcommand{\term}{{\it term}}
\newcommand{\pol}{{\it poly}}
\newcommand{\grob}{\hbox{\it Gröbner}}
\newcommand{\cset}{{\it Set}}
\newcommand{\cprop}{{\it Prop}}
\newcommand{\cfun}{\rightarrow}
\newcommand{\zerop}[1]{{\it zeroP}(#1)}
\newcommand{\lpol}{({\it list}\, \pol)}
\newcommand{\lterm}{({\it list}\, \term)}
\newcommand{\cano}{\mathcal{C}}
\newcommand{\olist}{\mathcal{O}}
\newcommand{\cforall}[3]{\forall #1\!\!:#2.\, #3}
\newcommand{\cforalln}[2]{\forall #1\!\!:#2}
\newcommand{\clambda}[3]{\lambda #1\!\!:#2.\, #3}
\newcommand{\cexists}[3]{\exists #1\!\!:#2.\, #3}
\newcommand{\cin}[2]{#1\,\,{\it in}\,#2}
\newcommand{\cdef}[3]{\hbox{{\bf Definition}$\,\, #1{\bf :}\, #2\,{\bf :=}\, \,#3$.}}
\newcommand{\cdefq}[2]{\hbox{{\bf Definition}$\,\, #1\,{\bf :=}\, \,#2$.}}
\newcommand{\cdefqn}[2]{\hbox{{\bf Definition}$\,\, #1\,{\bf :=}\, \,#2$}}
\newcommand{\cdeffn}[3]{\hbox{{\bf Definition}$\,\, #1{\bf :}\, #2\,{\bf :=}\,\,  #3$}}
\newcommand{\cdefp}[2]{\hbox{{\bf Definition}$\,\, #1\, {\bf :}\, \, #2$.}}
\newcommand{\cdefpn}[2]{\hbox{{\bf Definition}$\,\, #1\, {\bf :}\, \, #2$}}
\newcommand{\cdefrn}[2]{\hbox{{\bf Definition}$\,\, #1\, {\bf :}\, \, #2\,{\bf :=}$}}
\newcommand{\cthm}[2]{\hbox{{\bf Theorem}$\,\,  #1 {\bf :}\, #2$.}}
\newcommand{\cthmn}[2]{\hbox{{\bf Theorem}$\,\,  #1 {\bf :}\, #2$}}
\newcommand{\cthmrn}[1]{\hbox{{\bf Theorem}$\,\,  #1 {\bf :}$}}
\newcommand{\clem}[2]{\hbox{{\bf Lemma}$\,\, #1{\bf :}\quad  #2$.}}
\newcommand{\clemn}[2]{\hbox{{\bf Lemma}$\,\, #1{\bf :}\quad #2$}}
\newcommand{\cdefn}[2]{\hbox{{\bf Definition}$\,\, #1{\bf :}\quad  #2\, {\bf :=}$}}
\newcommand{\cfix}[3]{\hbox{{\bf Fixpoint}$\,\, #1\,{\bf [}#2{\bf ]:}\quad  #3\, {\bf :=}$}
}
\newcommand{\cfixn}[4]{\hbox{{\bf Fixpoint}$\,\, #1\,{\bf [}#2{\bf ]:}\quad  #3\, {\bf :=} #4$}
}
\newcommand{\caxiom}[2]{\hbox{{\bf Axiom}$\,\, #1{\bf :}\quad  #2$.}}
\newcommand{\caxiomn}[2]{\hbox{{\bf Axiom}$\,\, #1{\bf :}\quad  #2$}}
\newcommand{\cparam}[2]{\hbox{{\bf Parameter}$\,\, #1{\bf :}\quad  #2$.}}
\newcommand{\cparamn}[2]{\hbox{{\bf Parameter}$\,\, #1{\bf :}\quad  #2$}}
\newcommand{\ccas}[1]{\hbox{$\quad${\bf Cases}$\,\, #1\,\,{\bf of}$}}
\newcommand{\cend}{\hbox{$\quad${\bf end}}}
\newcommand{\cp}[1]{\hbox{#1.}}
\newcommand{\ccasf}[2]{\hbox{$\quad \ \, #1\,\, {\bf \Longrightarrow}\,\, #2$}}
\newcommand{\ccasi}[2]{\hbox{$\quad |\, #1\,\, {\bf \Longrightarrow}\,\, #2$}}
\newcommand{\clet}[2]{\hbox{$\quad${\bf let }$\, #1=#2\, {\bf in}$}}
\newcommand{\clin}[1]{\hbox{$\quad #1$}}
\newcommand{\cvar}[2]{\hbox{{\bf Variable}$\,\, #1{\bf :}\, #2$.}}
\newcommand{\chyp}[2]{\hbox{{\bf Hypothesis}$\,\, #1{\bf :}\, #2$.}}
\newcommand{\chypn}[2]{\hbox{{\bf Hypothesis}$\,\, #1{\bf :}\, #2$}}
\newcommand{\cind}[2]{\hbox{{\bf Inductive}$\,\, #1{\bf :}\, #2 :=$}}
\newcommand{\cindu}[3]{\hbox{{\bf Inductive}$\,\, #1[#2]{\bf :}\, #3 :=$}}
\newcommand{\cicn}[2]{\hbox{${\bf |}\quad #1\,{\bf :}\,#2$}}
\newcommand{\cicnf}[2]{\hbox{$\ \quad #1\,{\bf :}\,#2$}}
\newcommand{\cicnl}[2]{\hbox{${\bf |}\quad #1\,{\bf :}\,#2$.}}
\newcommand{\cicnfl}[2]{\hbox{$\ \quad #1\,{\bf :}\,#2$}}
\newcommand{\cic}[1]{\hbox{${\bf |}\quad #1$}}
\newcommand{\cicf}[1]{\hbox{$\ \quad #1$}}
\newcommand{\cicl}[1]{\hbox{${\bf |}\quad #1$.}}
\newcommand{\cicfl}[1]{\hbox{$\ \quad #1$.}}

\section{Definitions and validated properties}

\subsection{Floating point numbers}

An IEEE double precision floating  point number is built from 3 binary
fields:  the sign  (1  bit), the  fraction  (52 bits)  and the  biased
exponent  (11 bits).   Its  normal interpretation  $x$  as a  rational
number is given below.
$$
\begin{array}{r c l}
  \text{mantissa} & = & 1.\text{fraction} \\
  \text{exponent} & = & \text{biased exponent} - \text{bias} \\
               x  & = & (-1)^{\text{sign}} \times \text{mantissa} \times 2^{\text{exponent}}
\end{array}
$$


We  will use  in this  text  the more  general notation  $x= n  \times
\beta^{e}$ to define a floating point number of radix $\beta$ with two
integers, the  significand $n$ and  the amplitude $e$.   This relation
gives a  signification to  any pair  $(n, e)$.  We  will use  the name
floating point  number or {\bf float}  to represent such  a pair.  Two
floats $(n,  e)$ and $(n', e')$  are equivalent if they  have the same
value as real  numbers, we write $(n, e) \equiv  (n', e')$.  We extend
this notation to the represented real value and we may write $n \times
\beta^e  \equiv  (n  ,e)$.   We  will also  use  the  equivalence  for
expression whenever there is no ambiguity.  No order is defined on the
floats and inequalities are given implicitly on their interpretation.

Given a  pair $(n_{\max}, e_{\min}) \in  \N_*^2$, we say  that a float
$(n, e)$ is {\bf bounded} if and only if it satisfies
\begin{equation}
\label{eqn/bounded}
  |n| \le n_{\max}
~~~~~ \text{and} ~~~~~
  - e_{\min} \le e.
\end{equation}

We do  not use  the overflow bound  of $e_{\max}$.  The  IEEE standard
defines an overflow  when the rounded value of  the result exceeds the
largest  representable  number.   It  implies that  the  rounding  is
defined  on a data  type that  does not  have an  upper bound  for the
exponent.   All the  results of  this text  are true  provided neither
overflow  nor   exact  infinity  (division  by  0)   occurred  in  the
computation.  If this is not the case, infinity and {\em not a number}
quantities will proliferate in the results.

The bound for an IEEE standard inspired representation with $p$ digits
of mantissa and $r$ digits  of exponent is given Table~\ref{tab/ieee}. 
We will use the common fraction notation for the examples although the
integer  notation  is  used  for   the  proofs.   We  use  only  radix
independent generic IEEE standard inspired  bounds in the rest of this
text.   A  bounded  float is  {\bf  normal}  if  $|n| \times  \beta  >
n_{\max}$.  It is {\bf de-normal} if $|n| \times \beta \le n_{\max}$. A
de-normal float such that $e = -e_{\min}$ is called {\bf subnormal}. We
define the  {\bf ulp} value as  one unit in  the last place of  1 that
equals $\beta / (n_{\max} + 1)$.

\begin{table}
\caption{IEEE standard inspired bounds}
\label{tab/ieee}
\begin{center}
\begin{tabular}{| c | c | c | c |}\hline
           & Radix independent                            & \mc{2}{| c |}{$\beta = 2$}                     \\
           & (Generic)                                    & Single                & Double                 \\ \hline   
$p$        & Mantissa width                               & 24                    & 53                     \\
$n_{\max}$ & $\beta^p - 1$                                & 16~777~215            & 9~007~199~254~740~991  \\
$r$        & Exponent width                               & 8                     & 11                     \\
bias      & $\lceil  \beta^r / 2 \rceil  - 1$            & 127                   & 1023                   \\
$e_{\min}$ & $\lceil  \beta^r / 2 \rceil  + p - 3$        & 149                   & 1074                   \\
$e_{\max}$ & $\lfloor \beta^r / 2 \rfloor - p$            & 124                   & 971                    \\
ulp        & $\beta^{1 - p}$                              & $1.19 \times 10^{-7}$ & $2.22 \times 10^{-16}$ \\ \hline
\end{tabular}
\end{center}
\end{table}

\begin{theo}[FnormalizeCanonic and FcanonicUnique in Form]
  ~ Any bounded float has one unique equivalent float $(n, e)$ that is
  normal or subnormal.  This float  is the direct transcription of the
  number in machine with
  $$
  \begin{array}{r c l}
    n & = &  \text{mantissa} \times \beta^{p - 1}            \\
    e & = & (\text{biased exponent} - \text{biais}) - p + 1. \\
  \end{array}
  $$%
  It  is  later   referred  in  this  text  as   the  {\bf  canonical}
  representation. 
\end{theo}

The IEEE  standard describes four  rounding modes but the  rounding to
the nearest floating point number is the rounding mode used by default
in most  computers.  A  rounding is generally  defined as  a monotonous
projection onto the set of  the canonical floats.  Considering all the
floats rather than just the  canonical ones, we would rather define it
as a projection onto subsets of bounded floats.

For  any real  number  $x$, we  define  its bounded  floor $\lfloor  x
\rfloor$ as the class of the  maximum bounded floats smaller than $x$. 
It means  that a bounded float $(n,  e)$ is in $\lfloor  x \rfloor$ if
and only if, for all the bounded floats $(n', e')$ 
$$
(n', e') \le x ~~~~~ \Longrightarrow ~~~~~ (n', e') \le (n, e)
$$%
We define identically its bounded  ceil $\lceil x \rceil$ as the class
of  the  minimum  bounded  floats  larger than  $x$.   We  define  the
truncated subset $F(x)$ as follows.
$$
F(x) = \left\{\begin{array}{l c}
  \lfloor x \rfloor                     & x > 0 \\
  \lceil  x \rceil                      & x < 0 \\
  \{ (0, e), ~~ -e_{\min} \le e \}      & x = 0 \\
\end{array}\right.
$$

We also define the class of the bounded floats nearest to $x$ as $\Box
(x)$  with $(n,  e)$ is  in $\Box  (x)$ if  and only  if, for  all the
bounded floats $(n', e')$
$$
| (n, e) - x | \le | (n', e') - x|
$$

When $\Box  (x)$ contains two distinct equivalence  classes of bounded
floats,  we  define $\circ(x)$  as  the class  of  $\Box  (x)$ of  the
elements where  the canonical representation has an  even significand. 
In other cases,  $\circ(x) = \Box (x)$.  The  following theorem states
that this definition is appropriate.

\begin{theo}[*RoundedModeP in Fround and Closest]
  ~ The  relations rounding down  $\lfloor \cdot \rfloor$,  up $\lceil
  \cdot \rceil$, to zero $F  (\cdot)$ and to the nearest (even) $\circ
  (\cdot)$ define  each a total  monotonous projection from  the reals
  onto the  bounded floats compatible  with the interpretation  of the
  floats.
\end{theo}

We now give a few results on the rounding to the nearest.  The theorem
could be defined  on the canonical representation but  it is better to
prove it for any bounded float.

\begin{theo}[ClosestErrorBound in ClosestProp]
  \label{theo/Round/Error}%
  Let $x$ be any real number and $(n, e)$ any float of $\Box (x)$,
  $$
  |x - n \beta^e| \le \beta^e  / 2.
  $$
  That relation  is independent of the rounding being  even or not. 
  Even rounding provides that if both  $(n, e) \in \circ(x)$ and $|x -
  (n, e)| \equiv \beta^e / 2$ then $n$ is even.
\end{theo}

The following theorem gives a relation for every representation of the
rounded value and the error provided that the error is non zero and it
does have  a bounded representation.  The universal  quantifier is the
key to prove the theorems~\ref{theo/TS/Exp} and \ref{theo/M/Exp}.

\begin{theo}[RoundedModeErrorExpStrict in FRoundProp]
  \label{theo/Round/Error/Strict}
  Given an arbitrary rounding mode, let  $x$ be a real number and $(n,
  e)$  a  bounded  float  rounded  from  $x$, such  that  $x$  is  not
  represented by  $(n, e)$  ({\em ie.}  $x  \neq n \beta^e$).  For all
  bounded  float  $(n',  e')$ that  is  a  representation  of $x  -  n
  \beta^e$,
  $$
  e' < e.
  $$
  The bound  is tight as $(n', e')  = (1, e - 1)$  is an acceptable
  round-off error.
\end{theo}

The  result of  any implemented  operation, namely  the  addition, the
multiplication, the  division and the  square root extraction,  is the
rounded result  of the exact mathematical operation.   For example, if
$a \oplus b$ is  the machine addition then $a \oplus b  \in \circ (a +
b)$.  The machine  operation could  have  been defined  to return  the
canonical  float  in  $\circ(a  +  b)$.  We  preferred  to  define  an
acceptable implementation as a  function that return any bounded float
in $\circ(a + b)$.

When a theorem is independent of the tie breaking rule implemented, we
use the boxed symbols  $\boxplus$, $\boxminus$ and $\boxtimes$ for the
addition, the  subtraction and the  multiplication in the  hypothesis. 
Figure~\ref{fig:fpop}  presents  the   symbol  of  the  four  standard
floating point operators used in  this work.  All the figures use even
rounding as no other implementation exists.

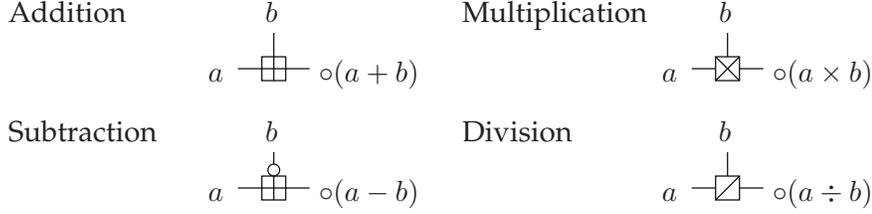
\figtex{fpop}{Standard  floating  point   operations  rounded  to  the
  nearest (even) value.}



\subsection{Exact operations}

We used for our proofs of the exact operation a result first published
in \cite{Ste74}. It was later  presented in \cite{Gol91} for the radix
2 notation.  This fact is true for any bounded representation with any
radix.

\begin{theo}[Sterbenz in Fprop]
  Given two  bounded floats $x$ and $y$  such that ${y \over  2} \le x
    \le 2  y$, the rational  $x - y$  can be represented by  a bounded
    float.
\end{theo}

We immediately prove  the next result from its  negation after setting
$\beta$ to be 2. We use it later in our proofs.

\begin{theo}[plusClosestLowerBound in Closest2Plus]
  Given two bounded floats radix 2, $x$ and $y$, such that $x + y \neq
  x \oplus y$, we know that $|x \oplus y| \ge \max (|x|, |y|) / 2$.
\end{theo}

M{\o}ller  first,  then  Knuth   proposed  the  common  exact  sum  of
Figure~\ref{fig:badd}-(a) involving  4 floating point  additions and 2
floating point subtractions on an IEEE compliant computer.  The result
is a pair of  floats $(a', b')$ such that $a' = a  \oplus b$ and $a' +
b' \equiv a  + b$.

\figtex{badd}{Exact sum}

We have proved this algorithm  correct (Theorem {\bf Knuth in TwoSum})
using  the proof given  by Priest  in~\cite{Pri91}.  Actually  we will
never use  directly the algorithm but  only the property  that the sum
and the  rounding error  can be represented.   This is  proved picking
bounded representations that have special properties:




\begin{theo}[plusExactExp in ClosestPlus]
  \label{theo/TS/Exp}
  Given two bounded floats $a =  (n_a, e_a)$ and $b = (n_b, e_b)$, for
  all $a' \in a \boxplus b$, we  can define $(n, e) \equiv a'$ and $b'
  = (n', e')$, bounded floats, such that
  $$
    a' + b' \equiv a + b
    ~~~~~ \text{and} ~~~~~
    \min (e_a, e_b) = e' \le e \le \max (e_a, e_b) + 1.
  $$
\end{theo}

This theorem uses the Theorem~\ref{theo/Round/Error} and Theorems {\bf
  errorBoundedPlus}, {\bf  plusExpBound} not presented  here to define
an    existing    representation    of    $a'$    and    $b'$.     The
theorem~\ref{theo/Round/Error/Strict} is used  to connect both results
since it is true for any  representation of $a'$ and $b'$. The special
case where $b' = 0$ is handled independently.


The last Theorem prompts the fact  that the addition is accurate up to
one unit  in the last  place. Such result  is needed to show  that the
numerical tools of the next Section actually do some work.

\begin{theo}[plusErrorBoundUlp in ClosestPlus]
  \label{theo/work}
  Given two bounded floats $a$ and  $b$, for all $a' \in a \boxplus b$
  such that $a' \not\equiv 0$, we know that
  $$
  |a + b - a'| \le       |a'|      {\ulp  \over 2}
               \le \max (|a|, |b|)  \ulp.
  $$%
  The last  inequality is true only  working radix 2. It  is proved in
  theorem plusErrorBound2 in file Closest2Plus.
\end{theo}

It has also  been proved by M{\o}ller and  presented in Knuth's second
edition that one can get a correct pair $(a', b')$ by an early exit of
the  exact  sum  provided  $|b|  \le |a|$.   The  conditional  sum  of
Figure~\ref{fig:badd}-(b) returns an exact pair $(a', b'') \equiv (a',
b')$  with only  2 floating  point  additions and  one floating  point
subtraction.

Knuth suggested that the early exit  is still valid if $a$ and $b$ are
both canonical  and share the same  exponent.  We prove  here a result
rephrased from \cite{Dau99a} that $a$  and $b$ just have to be bounded
and the exponent be in order.

\begin{theo}[ExtDekker in EFast2Sum]
  On a  radix 2 IEEE inspired  floating point unit,  given two bounded
  floats $a  = (n_a, e_a)$ and  $b = (n_b, e_b)$,  the conditional sum
  presented Figure~\ref{fig:badd}-(b) returns an exact pair $(a \oplus
  b, a + b - a \oplus b)$ provided
  $$e_b \le e_a.$$
\end{theo}

This  theorem is  proved only  for radix  two notations  as it  is not
necessarily  correct for other  radices.  An  example where  the early
exit is  not correct radix 10 is  given for $n_{\max} =  99$ with both
inputs  being  $9.9$.   The theorem  is  correct  for  radix 2  and  3
\cite{Knu73}. One can  check one last example radix  4: $n_{\max} = 3$
and both inputs are 3.

An exact multiplication  is also available.  It computes  a pair $(a',
b')$ such that  $a' = a \otimes b$ and  $a' + b' = a  \times b$ with 7
floating  point  multiplications, 5  floating  point  additions and  5
floating  point   subtractions.   These  operators   are  surveyed  in
\cite{Dau99a}.   Here are  some  properties obtained  from the  desired
specification of $a'$ and $b'$, not from the algorithm.  The condition
on the exponent is not necessary but it is sufficient to discard cases
of underflow.

\begin{theo}[multExactExpCan in ClosestMult]
  \label{theo/M/Exp}
  Given two bounded floats $a = (n_a, e_a)$ and $b = (n_b, e_b)$, with
  $e_a +  e_b \ge -e_{\min}  + p$, for  all $a'$ rounded to  a nearest
  float of $a \times b$, we  can define $(n, e) \equiv a'$ canonic and
  $b' = (n', e')$ bounded float, such that
  $$
    a' + b' \equiv a \times b
    ~~~~~ \text{and} ~~~~~
    e' = e - p.$$
\end{theo}

It was proved  that comparable error quantities, that  always fit in a
common  floating point  number, may  be defined  for the  division and
square root \cite{BohWalKorMat91}.


\subsection{Floating point expansions}

A floating point {\bf expansion} is defined as a finite sequence ${\bf
  x} = (x_0, x_1, \cdots , x_{n - 1})$ of floating point numbers.  The
value represented  is the  exact, not rounded,  sum of  its components
$\sum  x_i$.   The  length of  the  expansion  is  the number  of  its
components.


Any  component  of  an  expansion  may  be  equal  to  zero,  but  the
subsequence of  the non-zero components $x_i$ must  be non overlapping
and ordered  by magnitude.  The  non overlapping condition  means that
two components cannot have significant  bits with the same weight that
is specified for each $i > 0$ there exists a bounded float $(n, e)$
such that
$$
x_i \equiv (n, e) ~~~~~ \text{and} ~~~~~ |x_{i + 1}| < \beta^e.
$$



We have  already presented two examples in  \cite{Dau99a} showing that
any float  $(n, e)$  where $e  \ge -e_{\min}$ can  be expressed  as an
expansion.

As noted in the introduction,  previous works by Priest, Shewchuck and
ourselves have produced arithmetic  operators on expansions and useful
primitives for computational geometry.  In this process, we recognized
that the  operators for the addition and  the multiplication computing
the  least significant  digits  first do  not  produce length  optimal
results  but  tend  to  break  the  expansion in  a  large  number  of
components with  small significands. A compression routine  is used to
group  together the  components and  reduce the  length of  the result
expansion.



It is  possible to compute the arithmetic  operations most significant
digits first and compute directly components with a large significand.
The  common,  both restoring  and  non  restoring, division  operators
compute the components of the quotient like this.  As a drawback, some
of  the   components  may  overlap  slightly.   In   our  former  work
\cite{DauFin99a}, the sequence of the  quotient digits is cleaned by a
compression routine that produces an acceptable expansion.


We define {\bf pseudo-expansions}  as slightly overlapping expansions.
The condition on the subsequence of non zero components as that
$$|x_{i  + 1}|  \le \epsilon |x_i|.$$%
We do not give the value of $\epsilon < {1 \over 2}$ now as it will be
fixed  in the next  section considering  the numerical  algorithms. We
deduce that
$$|\sum_{j > i} x_j| \le {\epsilon \over 1 - \epsilon} |x_i|.$$
Conversely, given a sequence where 
$$|\sum_{j > i} x_j| \le \lambda |x_i|,$$
for all $i$, we deduce that
$$|x_{i  + 1}|  \le {\lambda \over 1 - \lambda} |x_i|.$$%




\section{Toolset for the addition, the multiplication and the division}

We combine in this Section  tools to create the arithmetic operations:
MQ,  PP, PQ  and $\Sigma_3$.  The tools  have been  implemented in  C. 
Among them, the numerical behavior of $\Sigma_3$ alone is difficult to
analyze as the  three other tools only sort  and produce components as
they are needed.


\subsection{Addition}

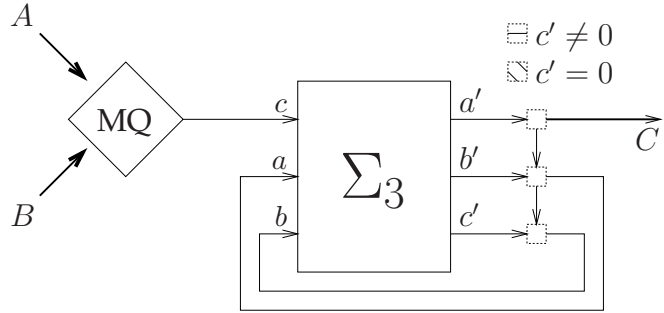
\figtex{addition}{Representation of the addition.}
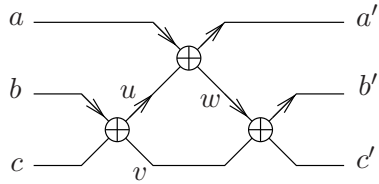
\figtex{insert3}{Definition of the $\Sigma_3$ operator.}

Figure~\ref{fig/addition}  represents  the  algorithm  implemented  to
compute  the  sum  of  pseudo-expansions  $A$ and  $B$.   We  use  two
primitive operators MQ (merge queue) and $\Sigma_3$ (insert and sum of
Figure~\ref{fig/insert3}).  Arrows represent streams of floating point
number flowing  most significant digits first.  The  components of the
result are in the  stream named $C$ in Figure~\ref{fig/addition}.  The
output  is  only a  pseudo-expansion  since  the  $a'$ are  sorted  by
decreasing  order of  magnitude but  they may  partially  overlap each
other.

MQ is an  operator fired only if it  has a value on both  inputs or if
the flow  of one input  is finished  and it has  a value on  the other
input. The  largest input,  in magnitude, is  then transmitted  to the
output.  Cases of tie  are not  critical. The  MQ operator  merges two
flows sorted by magnitude and produces a single flow once again sorted
by magnitude.  The following theorem  states that this flow  of values
can be represented  by a flow of floats to be  fed into the $\Sigma_3$
operator.

\begin{theo}[IsRleExpRevIsExp in FexpAdd]
  Given a list of floats $(x_i)_{i \in L}$ sorted by magnitude, we can
  define the bounded floats $(n_i, e_i)$ equivalent to the $x_i$s such
  that the list $(e_i)_{i \in L}$ is sorted.
\end{theo}

We deduce the correct behavior  of the operation by induction from the
following Theorem.   The property on  the input floating  point number
$c$ is inherited from the fact  that the two input streams $A$ and $B$
are pseudo-expansions merged by the MQ operator by order of magnitude.

\begin{theo}[bound3Sum and exp3Sum in ThreeSum2]
  \label{theo/sum3}
  Let $a = (n_a,  e_a)$, $b = (n_b, e_b)$ and $c  = (n_c, e_c)$ be the
  bounded     inputs    of     the     $\Sigma_3$    operator     (see
  Figure~\ref{fig/insert3}).  Provided that $e_a  \ge e_b \ge e_c$ the
  $\Sigma_3$   operator  returns  three   numbers  $a'$,   $b'$,  $c'$
  represented  by $(n_a',  e_a')$, $(n_b',  e_b')$ and  $(n_c', e_c')$
  such that $e_a' \ge e_b' \ge e_c'  = e_c$ and finaly either $c' = 0$
  or
  $$|b' + c'|   \le  3 \ulp                   |a'|,$$
  $$\beta^{e_c} \le {3 \ulp^2 \over 2 - \ulp} |a'|.$$
  
  If $a = 0$ or $|b| + n_{\max} \beta^{e_c} \le n_{\max} \beta^{e_a}$,
  all the exact additions can use the early exit.
\end{theo}

If $c'$  equals to zero, $a'$  and $b'$ are not  relevant enough. They
remain in the operator as $a$  and $b$ for the next iteration.  On the
opposite, if $c'$  is not equal to zero, $a'$ is  relevant.  It is one
component of  the result, $b'$ and  $c'$ are kept in  the operator for
the next iteration.

\paragraph{\em Proof sketch:}

We first define $u$, $v$, $a'$,  $b'$ and $c'$ as they are computed in
Figure~\ref{fig/insert3}. We are interested in the case where $c' \neq
0$. It implies that $w \neq 0$. We bound
$$
\begin{array}{r c l}
  |w| & \le & {\ulp \over 2} |a'| \\
  |v| & \le & \ulp \max (|u|, |a|)
\end{array}
$$
Since $w \neq 0$, we know that 
$$|a'| \ge {\max (|u|, |a|) \over 2}.$$
The first bound follows since
$$|b' + c'| = |v + w| \le |v| + |w|.$$

We continue for the second bound
$$
3 \ulp |a'| \ge |b' + c'|
            \ge |b'| - |c'|
            \ge \left( 1 - {\ulp \over 2} \right)|b'|.
$$
We conclude with  the representation  of Theorem~\ref{theo/TS/Exp}
$$
\beta^{e_c} = \beta^{e_c'}
            \le |c'|
            \le {\ulp \over 2} |b'|
            \le {3 \ulp^2 \over 2 - \ulp} |a'|.
$$

\begin{flushright}
$\Box$
\end{flushright}

The last theorem does the induction over time. It is written from both
statement and proof of the theorem in Coq.

\begin{theo}[FexpAdd in FexpAdd]
  Given a list $(n_i, e_i)_{i \in  L}$ of $\#L$ bounded floats so that
  $(e_i)_{i   \in   L}$  is   sorted   the   addition  operator   (see
  Figure~\ref{fig/addition}) produces  a list $(c_i)_{i \in  L}$ of at
  most $(\#L + 1)$ floats with
  $$
  |c_{i + 1}| \le \epsilon |c_i|
  ~~~~~ \text{with} ~~~~~
  \epsilon \le {6 \#L + 6 \over n_{\max} - 1 - 6\#L}
  $$
  under the condition that $\epsilon < 1$.
\end{theo}

\paragraph{\em Proof sketch:}

We have  to show that the  input conditions are met  at each iteration
using  pre-conditions  and  post-conditions.   In the  same  time,  we
establish that
$$|\sum_{j > i} c_j| \le 3(1 + 2 \#L) \ulp |c_i|,$$
and
$$|c_{i  + 1}|       \le 6(1 + 2 \#L) \ulp |c_i|.$$%
Being more careful, we proved the tighter bound in Coq.

\begin{flushright}
$\Box$
\end{flushright}


\subsection{Multiplication}

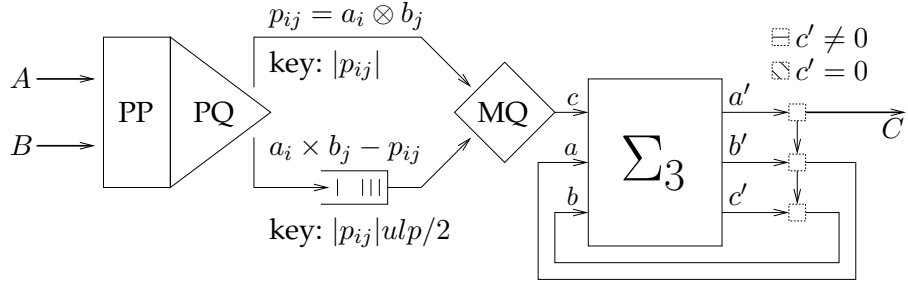
\figtex{multiplication3}{Representation of the multiplication.}

The  multiplication  of  the   expansions  $A$  (size  $n_A$)  by  the
expansions  $B$ (size $n_B$)  generates the  $n_A \times  n_B$ partial
products and  computes their sum. The  problem is to  sort the partial
products  $a_{i_j} \times  b_j$  knowing that  the  lists $(a_i)$  and
$(b_j)$ are  sorted by  magnitude. It  is related to  $X +  Y$ sorting
\cite{HarPaySavStr75,SteStr94}.        Figure~\ref{fig/multiplication3}
represents  the  algorithm  implemented.   We  use  new  and  extended
primitives: PP, a  partial product generator, PQ and  a priority queue
extending the MQ.


Working with pseudo-expansions, we have  to make sure that the list of
the  partial products  is sorted  by magnitude.   As $b_j$  arrives at
position  $j$,  the PP  tool  initializes the  $i_j$  index  to 0  and
computes $a_{i_j} \otimes b_j$.  Its  magnitude is inserted at the end
of PQ that is updated in $\lfloor \log j \rfloor$ operations. When the
partial  product $a_{i_j}  \otimes  b_j$  is output  by  PQ, the  tool
increments  $i_j$.  The  generation  is frozen  until both  components
$a_{i_j}$  and  $b_{j +  1}$  are  known.   Then the  cell  containing
$|a_{i_j} \otimes b_j|$  in PQ is updated in  $\lfloor \log j \rfloor$
operation and the tools can generate the next partial product.


Numbers  flow by  pair since  $a_i \times  b_j$ produces  two floating
point  numbers. The  upper part  $p_{ij} =  a_i \otimes  b_j$  is used
immediately. The  lower part $a_i \times  b_j - p_{ij}$ is  put into a
waiting  queue.   The  two  streams  are merged  together  by  the  MQ
operators. It is fired as long as it has an input on both entries.  It
does not compare the values but  the keys that are associated with the
floats.   The operator does  not produce  a list  of floats  sorted by
magnitude but directly a list sorted by amplitude as we use the second
part of theorem~\ref{theo/M/Exp}.

As $a_i \times  b_j - p_{ij}$ cannot be  selected before $p_{ij}$, its
insertion  into  the  waiting queue  is  not  in  the critical  path.  
Instruction reordering will allow these operation to be hidden in dead
cycles of the $\Sigma_3$ operator.   The waiting list cannot hold more
than $2n_B$ elements since the bound on $\epsilon$ is a few ulps.




\subsection{Division}


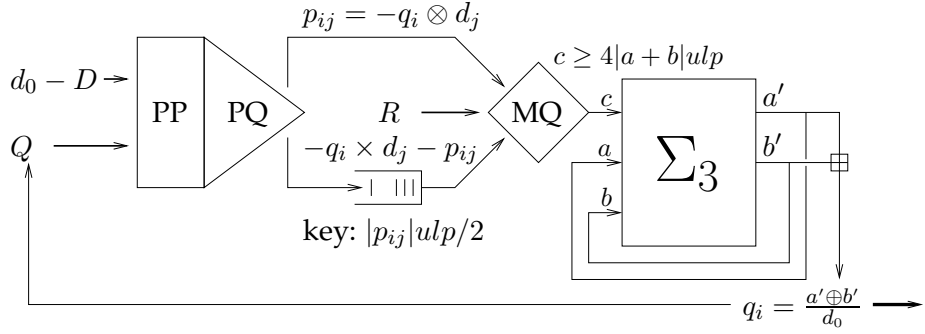
\figtex{division3}{Representation of the non-restoring division.}

Figure~\ref{fig/division3}  represents  the division  of  $R$ by  $D$,
pseudo-expansions. We want to compute $Q$  such that $Q = R / D$, {\em
  ie.}  $R - QD = 0$. Each  component $q_i$ of $Q$ is deduced from the
previous ones by computing
$$R_i = R - \sum_{j < i} q_j \times D,$$%
before evaluation $q_i \approx R_i /  D$. We will see later that $d_0$
is handled separately, so our scheme computes $Q \times (d_0 - D)$.

The  goal is  to ensure  that the  values $c$  used as  inputs  of the
$\Sigma_3$ operator  are sorted by  amplitude.  The PP-PQ-MQ  chain is
close   to   the   one    used   for   the   multiplication   in   the
Figure~\ref{fig/multiplication3}. The  generation is frozen  only when
$d_{j + 1}$ is not known.

The test  on $q_{i_j}$ is replaced  by a new  condition: the operation
stops before it may insert a value $c$  such that $c < 4 |a + b| \ulp$. 
A new  approximate quotient  digit $q_i$ is  then guessed from  a fair
most significant  component $d_0$ of the  divisor $D$ and  a fair most
significant component $a' \oplus b'$.




The following  theorem states that one  division step is  very robust. 
For a reasonable value of $\epsilon$, it guarantees that the remainder
decreases almost linearly.

\begin{theo}[DivConv in FexpDiv]
  \label{theo/Div/Converge}
  Let $W$ and $D$ be two non zero real numbers approximated by $w$ and
  $d$ with  a relative error  bound $\epsilon <  1$ and let $q$  be an
  approximation of $w / d$ with the same relative error bound.
  $$
  |W - q D | \le \epsilon {3 + \epsilon \over 1 - \epsilon} |W|
  $$
\end{theo}

To reduce $\epsilon$, we replace the initial values $d_0$ and $d_1$ by
$d_0 \oplus d_1$  and $d_0 + d_1 - d_0 \oplus  d_1$. As a consequence,
the worst  error on the Theorem~\ref{theo/Div/Converge} is  the one on
the  remainder and  it  is bounded  by  construction by  4 ulps.  This
guarantees with the  loop condition that any term  $q_i \times d_j$ is
smaller than  all the  terms already produced  by the  partial product
generation chain except  $q_i \times d_0$. This last  value is handled
separately as $a'$ and  $b'$ are replaced by $a' + b'  - q_i d_0$. The
loop condition also ensures that $c' = 0$.







\section*{Conclusion}

We  have presented  a new  adaptable numeric  core inspired  both from
floating point expansions and from on-line arithmetic.  Our choice was
to  present only  the arithmetic  core as  we have  presented  in past
publications  how a  core can  be used  efficiently to  compute matrix
determinants for example \cite{DauFin99a}.  Building a general purpose
runtime environment raises question  in areas like parallelism (demand
driven {\em vs.}  data flow...)  and in compiler techniques (efficiency
of  object oriented  code generated  by existing  compilers...)  among
others.
  
The numeric core  is cut down to four  tools.  The $\Sigma_3$ operator
that  contains  many arithmetic  operations  is  proved correct.   The
proofs  have been  validated  by the  Coq  assistant.  Developing  the
proofs,  we have  formally proved  many result  long published  in the
literature and  we have  extended a  few of them.   This work  may let
users
\begin{itemize}
\item Develop  application specific  adaptable libraries based  on the
  toolset.
\item  Easily write  new formal  proofs based  on the  growing  set of
  sensible validated facts.
\end{itemize}

The set  of the validated facts  is now sufficient to  shield the user
from   the  difficult   implementation  details   of   floating  point
arithmetic.  The proof  of the key Theorem on  the $\Sigma_3$ operator
does not  use any low  level result. It  just uses the many  lower and
upper bounds defined in the Theorems of Section~1.

\small


\appendix
\section{A brief overview of Coq}

In this appendix,  we give a quick overview of the  Coq system.  For a
more complete introduction, we refer the reader to~\cite{HueKahPau97}.

Coq is a generic prover based on type-theory. In this system,
users can define new objects {\it and} prove properties that
derive logically from these definitions. Objects in Coq are
typed and functions are first-class objects using a Lisp-like
notation. The system is distributed with standard libraires that
define types like {\tt nat}, {\tt Z} and {\tt R} which correspond to
the natural numbers, the relative numbers and the reals respectively. 
To give an example, the addition for natural number is represented by 
the function {\tt plus}
whose type is {\tt nat $\rightarrow$ nat $\rightarrow$ nat}.
A function {\tt plus3} that does the sum of 3 natural numbers is
defined by the following command:
{\small \begin{verbatim}
Definition plus3 := [a,b,c:nat] (plus a (plus b c)).
\end{verbatim}}
\noindent 
Arguments between square brackets provide the parameters of the function.
The type of {\tt plus3} is the expected one: {\tt nat $\rightarrow$  nat $\rightarrow$ 
nat $\rightarrow$ nat}.

In our formalization, we first define the type {\tt float} that
represents the record composed of the mantissa and the exponent
by the command:
{\small \begin{verbatim}
Record float: Set := Float {Fnum: Z; Fexp: Z}.
\end{verbatim}}
\noindent
This command creates a new type {\tt float}, a constructor {\tt Float}
of type {\tt Z $\rightarrow$  Z $\rightarrow$ float} and two
destructors {\tt Fnum} of type {\tt  float $\rightarrow$ Z} and
{\tt Fexp} of type {\tt  float $\rightarrow$ Z}. So for example,
we can define the function {\tt Fzero} that takes an object of
type {\tt float} and returns an object of same exponent but
with the mantissa set to zero:
{\small \begin{verbatim}
Definition Fzero := [p:float](Float 0 (Fexp p)).
\end{verbatim}}
\noindent
In a similar way, we define the notion of bound. It contains
two integers: one for the mantissa and one for the exponent:
{\small \begin{verbatim}
Record Fbound: Set := Bound {vNum: nat;dExp: nat}.
\end{verbatim}}
\noindent
In order to prove properties that are independent of a particular
bound, we add the command:
{\small \begin{verbatim}
Variable b:Fbound.
\end{verbatim}}
\noindent
With this arbitrary bound, the notion of being bounded is defined as:
{\small \begin{verbatim}
Definition Fbounded := [p:float]
    ((Zle (Zopp (vNum b)) (Fnum p))  /\ (Zle (Fnum p) (vNum b))) /\
    (Zle (Zopp (dExp b)) (Fexp p)).
\end{verbatim}}
\noindent
A pretty-printed version of this definition could be:\\
\vbox{
\vskip5pt
\hbox{{\tt Definition Fbounded := $\lambda$p:float.\,}}
\hbox{ {\tt -(vNum b) $\le$ (Fnum p) $\le$ (vNum b) $\land$ -(dExp b) $\le$ (Fexp p).}}
\vskip5pt
}\\
Now that we have these definitions we can prove a very simple
first theorem:
{\small \begin{verbatim}
Theorem BoundedZero: (p:float) (Fbounded p) ->(Fbounded (Fzero p)).
\end{verbatim}}
\noindent
A pretty-printed version of this theorem could be:\\
\vbox{
\vskip5pt
\hbox{{\tt Theorem BoundedZero: $\forall$p:float.\, (Fbounded p) $\Rightarrow$ (Fbounded (Fzero p)).}}
\vskip5pt
}\\
When the previous command is received by the system, it enters 
the proof mode. The statement of the theorem is put om the top
of the goal stack. By applying a command called {\it tactic} we
replace the goal at the top of the stack by a (possibly empty)
list of subgoals. The proof is finished when the stack is empty.
Our initial stack has only one goal. Each goal contains a list
of hypothesis and a conclusion separated by a bar:
{\small \begin{verbatim}
1 subgoal
  
  ============================
   (p:float)(Fbounded p)->(Fbounded (Fzero p))
\end{verbatim}}
\noindent
The first step is to introduce the hypothesis. For this,
we apply the tactic {\tt Intros}.
{\small \begin{verbatim}
Intros p H.
\end{verbatim}}
\noindent
We are now reduce to prove {\tt (Fbounded (Fzero p))} in
the contect where {\tt p} is a {\tt float} and {\tt H} is a 
proof that p is bounded:
{\small \begin{verbatim}
1 subgoal

  p : float
  H : (Fbounded p)
  ============================
   (Fbounded (Fzero p))
\end{verbatim}}
\noindent
Next we expand the definition of {\tt Fbounded} in the goal. For this,
we use the tactic {\tt Red} and get the new goal:
{\small \begin{verbatim}
1 subgoal
  
  p : float
  H : (Fbounded p)
  ============================
   ((Zle (Zopp (vNum b)) (Fnum (Fzero p)) /\ 
    (Zle (Fnum (Fzero p)) (vNum b))) /\
    (Zle (Zopp (dExp b)) (Fexp (Fzero p)))
\end{verbatim}}
\noindent
This goal can be simplified using the definition of
{\tt Fzero} by the tactic {\tt Simpl}:
{\small \begin{verbatim}
1 subgoal
  
  p : float
  H : (Fbounded p)
  ============================
   ((Zle (Zopp (vNum b)) 0) /\ (Zle 0 (vNum b))) /\
    (Zle (Zopp (dExp b)) (Fexp p))
\end{verbatim}}
\noindent
We now need to break the conjunctions into separate
subgoals. This is done by the tactic {\tt Split}. As
the conjunction is nested, this tactic needs to be
applied repeatedly: 
{\small \begin{verbatim}
Repeat Split.
\end{verbatim}}
\noindent
We get the expected three subgoals.
{\small \begin{verbatim}
3 subgoals
  
  p : float
  H : (Fbounded p)
  ============================
   (Zle (Zopp (vNum b)) 0) 

subgoal 2 is:
   (Zle 0 (vNum b))

subgoal 3 is:
   (Zle (Zopp (dExp b)) (Fexp p))
\end{verbatim}}
\noindent
The first goal is simple enough to be proved automatically
by the following tactic {\tt Intuition}. We are left with two subgoals.
{\small \begin{verbatim}
2 subgoals
  
  p : float
  H : (Fbounded p)
  ============================
   (Zle 0 (vNum b))

subgoal 2 is:
   (Zle (Zopp (dExp b)) (Fexp p))
\end{verbatim}}
\noindent
Applying the same tactic to these two subgoals ends the proof.
The final proof script looks like:
{\small \begin{verbatim}
Theorem BoundedZero: (p:float) (Fbounded p) ->(Fbounded (Fzero p)).
Intros p H.
Red.
Simpl.
Repeat Split.
Intuition.
Intuition.
Intuition.
Qed.
\end{verbatim}}
\noindent
Using the composition of tactics ";", it can be shorten to:
{\small \begin{verbatim}
Theorem BoundedZero: (p:float) (Fbounded p) ->(Fbounded (Fzero p)).
Intros p H; Red; Simpl; Repeat Split; Intuition.
Qed.
\end{verbatim}}

\end{document}